# A Framework for Explainable Text Classification in Legal Document Review

Christian J. Mahoney
e-Discovery
Cleary Gottlieb Steen & Hamilton LLP
Washington DC, USA
cmahoney@cgsh.com

Dr. Jianping Zhang
Data & Technology
Ankura Consulting Group, LLC
Washington DC, USA
jianping.zhang@ankura.com

Nathaniel Huber-Fliflet
Data & Technology
Ankura Consulting Group, LLC
Washington DC, USA
nathaniel.huber-fliflet@ankura.com

Peter Gronvall
Data & Technology
Ankura Consulting Group, LLC
Washington DC, USA
peter.gronvall@ankura.com

Dr. Haozhen Zhao
Data & Technology
Ankura Consulting Group, LLC
Washington DC, USA
haozhen.zhao@ankura.com

*Abstract*— Companies regularly spend millions of dollars producing electronically-stored documents in legal matters. Over the past two decades, attorneys have been using a variety of technologies to conduct this exercise, and most recently, parties on both sides of the 'legal aisle' are accepting the use of machine learning techniques like text classification to cull massive volumes of data and to identify responsive documents for use in these matters. While text classification is regularly used to reduce the discovery costs in legal matters, text classification also faces a peculiar perception challenge: amongst lawyers, this technology is sometimes looked upon as a "black box." Put simply, very little information is provided for attorneys to understand why documents are classified as responsive. In recent years, a group of AI and Machine Learning researchers have been actively researching Explainable AI. In an explainable AI system, actions or decisions are human understandable. In legal 'document review' scenarios, a document can be identified as responsive, as long as one or more of the text snippets (small passages of text) in a document are deemed responsive. In these scenarios, if text classification can be used to locate these responsive snippets, then attorneys could easily evaluate the model's document classification decision. When deployed with defined and explainable results, text classification can drastically enhance the overall quality and speed of the document review process by reducing the time it takes to review documents. Moreover, explainable predictive coding provides lawyers with greater confidence in the results of that supervised learning task.

This paper describes a framework for explainable text classification as a valuable tool in legal services: for enhancing the quality and efficiency of legal document review and for assisting in locating responsive snippets within responsive documents. This framework has been implemented in our legal analytics product, which has been used in hundreds of legal matters. We also report our experimental results using the data from an actual legal matter that used this type of document review.

## I. INTRODUCTION

Big data in the legal domain has created serious business challenges. In modern litigation in the United States, attorneys often face an overwhelming number of documents that must be reviewed and produced over the course of a legal matter. The costs involved in manually reviewing these documents have grown dramatically, because for all practical purposes, the volumes of data implicated in modern legal matters are insurmountably large. As a result of this volume problem, the document review process can require an extraordinary dedication of resources: companies routinely spend millions of dollars sifting through and producing responsive electronically stored documents in legal matters [1].

For more than ten years, attorneys have been using machine learning techniques like text classification to more efficiently cull massive volumes of data to identify responsive information. In the legal domain, text classification is typically referred to as predictive coding or technology assisted review (TAR). Predictive coding applies a supervised machine learning algorithm to build a predictive model that automatically classifies documents into attorney-defined categories. These categories are often related to identification of "responsive" documents, which are materials that fall within the scope of some 'compulsory process' request, including discovery requests for production, a subpoena, or an investigative demand. This classification process is also used by attorneys outside of responding to production requests including for identification of interesting documents in internal investigations or litigations and for identification and protection of attorney client privilege and confidential information. While predictive coding is regularly used to reduce the discovery costs of legal matters, it also faces a perception challenge: amongst lawyers, this technology is sometimes looked upon as a "black box." Put simply, the process typically provides an ultimate classification of a document but provides little to no information to attorneys as

to why an individual document received a specific classification such as "responsive."

This study undertakes research to address a still not-widely studied component of the predictive coding process: explaining why documents are classified as responsive. Attorneys typically want to know why certain documents were determined to be responsive by a predictive model, but sometimes those answers are not obvious or easily divined. In many instances, for example, a predictive model could be inaccurate, resulting in flagging nonresponsive documents as highly likely to be responsive. This can be confusing for an attorney, especially if the text content of the document does not appear to contain obvious responsive content. An attorney with extensive knowledge of the training documents can make an educated guess as to why a document was classified as responsive, but it can be difficult to pinpoint exactly what text in a document heavily influenced the decision. The focus of our research here was to develop a framework of approaches that can target and examine the document text identified by the predictive model that was used to make the classification decision.

Explainable Artificial Intelligence, as a research topic, has attracted interest from the Artificial Intelligence (AI) community since the 1970s, e.g. medical expert system MYCIN [2]. In recent years, DARPA proposed a new research direction for furthering research into Explainable AI (XAI) [3]. In XAI systems, actions or decisions are human understandable – "machines understand the context and environment in which they operate, and over time build underlying explanatory models that allow them to characterize real world phenomena." Similarly, in an explainable machine learning system, predictions or classifications generated from a predictive model are explainable and human understandable. Interpreting the decision of the machine is more important now than ever because, increasingly, machine learning systems are being used in human decision-making applications and machine learning algorithms are becoming more complex.

Understanding a model's classification decision is challenging in text classification because the model considers an array of factors during the decision-making process, including word volume within text-based documents, and the volume and diversity of words established during the text classification process. In the legal domain, where documents can range from one-page emails to spreadsheets that are thousands of pages long, the complexity of the models creates challenges for attorneys to pinpoint where the classification decision was made within a document.

In legal document review, a document is considered responsive when one or more of its text snippets are responsive and contain relevant information. This is not always true for many other text classification tasks. For example, in topic classification, when a document is classified to a topic, the entire document may talk about the topic. For the purposes of this paper, we considered a text snippet to be a small passage of words within a document usually ranging from 50 to 200 words. Therefore, if predictive coding could locate these responsive snippets, attorneys could easily evaluate the model's document classification decision. In this scenario, there would be greater clarity as to why the model made its classification decision.

In addition to creating an explainable result, explainable predictive coding could enhance the overall document review process by reducing the time to review documents. Consider a scenario where attorneys – encumbered with a massive volume of potentially relevant documents – are presented with the option of only needing to focus on designated, responsive snippets of text within certain documents, and not the entire text of a document. To any attorney, this option would significantly speed up the review process. Explainable predictive coding has a perfect application in the legal document review process. It could also help improve investigative scenarios by quickly pinpointing potentially sensitive responsive content and enable a quick summary review of a small number of high scoring document snippets. Quickly understanding the content in a data set equips attorneys with the information they need to effectively represent their clients.

Within real-world legal matters, the Ankura authors of this article developed an analytics platform for legal document review scenarios. The platform's core component is Predict, a text classification application. In addition to text classification, the analytics platform contains other analytics components, such as near-duplicate document detection, image classification, document and image clustering, and sentiment analysis. In the last three years, the platform has been used in more than 130 legal matters and ingested and processed more than five hundred million documents.

This paper describes a framework of approaches for explainable text classification for legal document review (explainable predictive coding). This framework provides an effective methodology that lawyers can use to locate responsive snippets within a responsive document. The explainable predictive coding framework was implemented as a function in the predictive coding component of Ankura's analytics platform and applied to a real legal matter. We report our experimental results using the data from a real legal matter that has since concluded. The rest of the paper is organized as follows: Section II discusses previous work in explainable text classification. We introduce our explainable predictive coding framework in Section III and report the experimental results in Section IV. In Section V we discuss an additional observation and conclude in Section VI.

## II. PREVIOUS WORK IN EXPLAINABLE TEXT CLASSIFICATION

Interpreting the results of machine learning models has drawn the interest from both academic and industry experts, due to the increasing role that these models play in assisting humans with making decisions, based upon their results. Research in Explainable AI has focused on two main approaches of explainable machine learning: model-based explanations and prediction-based explanations. In a model-based approach, certain types of machine learning models, such as decision-tree models or 'if-then' rules-based models, are inherently easy to interpret (or "explain") using a human's point of view. Within this approach, the ultimate goal is to create machine learning models that are either based on interpretable models or ones that can be approximated or reduced into interpretable model components. Complex models, such as deep learning models like multilayer neural network models, non-linear SVM, or ensemble models, are not directly human understandable and require implementing a more sophisticated approach to interpret the decision. For example, an adversarial training scheme was proposed in [4] to trace the eventual outcomes back to interpretable representations consisting of influential neurons in the Deep Neural Networks (DNNs) model.

Another explainable machine learning approach is prediction-based and creates "explanations" for each individual prediction generated by a complex model: i.e. to explain the outcome of the model. Generally, a prediction-based explanation approach provides an explanation as a vector with real value weights, each for an independent variable (feature), indicating the extent to which it contributes to the classification. This approach is not ideal for text classification, due to the high dimensionality in the feature space. In text classification, a document belongs to a category, most likely because some passages of the text in the document support the classification. Therefore, a small portion of the document text is often used as evidence to justify the classification decision in text classification.

Predictive coding is an application of text classification used to identify documents that fall within the scope of a legal matter. Text data (documents) are often represented using the bag-of-words approach and characterized with tens of thousands of variables (words or phrases). Due to the high dimensionality, understanding the text classification models' decisions is very difficult, which creates an interesting research opportunity for this group of collaborators [5].

Recent research found that a prediction-based approach is often used to identify snippets of text as an explanation for the classification of a document. The underlying thinking is, a text snippet would provide an observer of the model's results with a brief 'rationale' as to why the document was identified as responsive to the underlying search. [6]. From a machine learning training perspective, annotated rationales provide more effective labeled input due to their targeted evidence relating to the relevance of the decision. Several research examples show that rationales improve model performance. Zaidan, et al. [6] proposed a machine learning method to use annotated rationales in documents to boost text classification performance. In their method, the labeled documents, together with human-annotated text snippets, were used as training data to build a text classification model using SVM. Experiment results showed that classification performance significantly improved with annotated rationales over the baseline SVM variants using an document's entire text. In [5], it is showed that classification models trained using only rationales and sampled not responsive text snippets, performed significantly better than models trained using entire responsive and not responsive documents when classifying responsive text snippets from not responsive ones. Zhang et. al. exploited rationales in augmenting convolutional neural networks models for text classification, by boosting the rationales' contribution to the aggregated document representation and found that the augmented model consistently outperforms strong baselines [7].

An essential part of prediction-based explainable machine learning is to generate the rationales for text classification that serve as document-level explanations for the predictive model's performance. Zaidan et. al is among the first to model human annotators to identify contextual rationales in document [8]. They used a generative approach and the model was trained on human-annotated rationales. Lei et al. [9] proposed a neural network approach to generate rationales for text classifications. Their approach combined two components: a rationale generator and a rationale encoder, which were trained to operate together. The generator specified a distribution over text fragments (text snippets) as candidate rationales and the encoder decided the classifications of candidate rationales established by the generator. The proposed approach was evaluated on multi-aspect sentiment analysis against manually annotated test cases and the results showed that their approach outperformed the baseline by a significant margin. Chhatwal et. al. used models trained using either entire documents or annotated rationales to identify rationales within overlapping text passages and demonstrated that models created with the full document text can successfully identify rationales at close to 50% recall by only reviewing the top two ranked passages [5].

In addition to using rationales directly from the target document text to explain the prediction, there are approaches that derive rationales from other sources. In [10], Martens and Provost described a method in which the explanation of a document classification was a minimal set of the most relevant words, such that removing all the words in the set from the document would change the classification of the document. An algorithm for finding this minimal set of words was presented and they conducted case studies demonstrating the value of the method using two sets of document corpora. In the popular LIME (Local Interpretable Model-agnostic Explanations) tool, predictions of a black box text classifier are explained by creating an interpretable model that provides explanations in the form of positive and negative class words that are most relevant to the individual predictions [11]. Other related research efforts include deriving precise attribute (or aspect) value predictions to serve as the explanation of the predictions [12].

## III. AN EXPLAINABLE PREDICTIVE CODING FRAMEWORK

The main goal of this proposed explainable predictive coding framework is to provide additional information (explanations) about a predictive model's labeling decision and to help attorneys more effectively and efficiently identify responsive documents during legal document review. As with other explainable text classification approaches, we use the prediction-based approach instead of the model-based approach. Additionally, we are only interested in generating explanations for responsive documents, therefore we focus on documents identified as responsive.

An explanation of a responsive document is a text snippet, referred to as rationale, in the responsive document. Figure 1 illustrates the architecture of the framework. A document classification model is first learned from a set of training documents using a machine learning algorithm. The document model is then applied to classify unlabeled documents into responsive and not responsive documents. In the next step, the snippet detection component is applied to all detected responsive documents to try to identify one or more rationales for each responsive document using the document model. In addition to rationales, a set of responsive tokens is also identified and these tokens may occur anywhere in the document. The model is used to find the tokens that support the classification. These responsive tokens can also be used to identify rationales.

Explainable predictive coding sets out to build a method to estimate the following probability:

$$Pr(r = Rationale \mid x, y = Responsive) \cdot Pr(y = Responsive \mid x),$$

where $x$ is a document, $y$ is the model-labeled designation of the document (for example, 'responsive' or 'not responsive') and $r$ is a text snippet from $x$.

In our proposed framework, the document classification model was used to also identify rationales. In [5], we compared the performance of document classification models and rationale classification models and the performance of document models are comparable with the performance of rationale models. A rationale model is trained using labeled rationales, but labeled rationales are not typically practical to create in most real legal matters. In this paper, we propose a novel approach to identify rationales using the document model.

Table 1 outlines the process for identifying rationales of responsive documents. In the following subsections, we describe the framework's three components and approaches to identify rationales based on individual components or integrating the three components together.

Table 1: The Process for Identifying Rationales

| |
|---|
| 1. Train a document model. |
| 2. Use the document model to identify responsive documents. |
| 3. Break each responsive document into a set of overlapping text snippets with n words. |

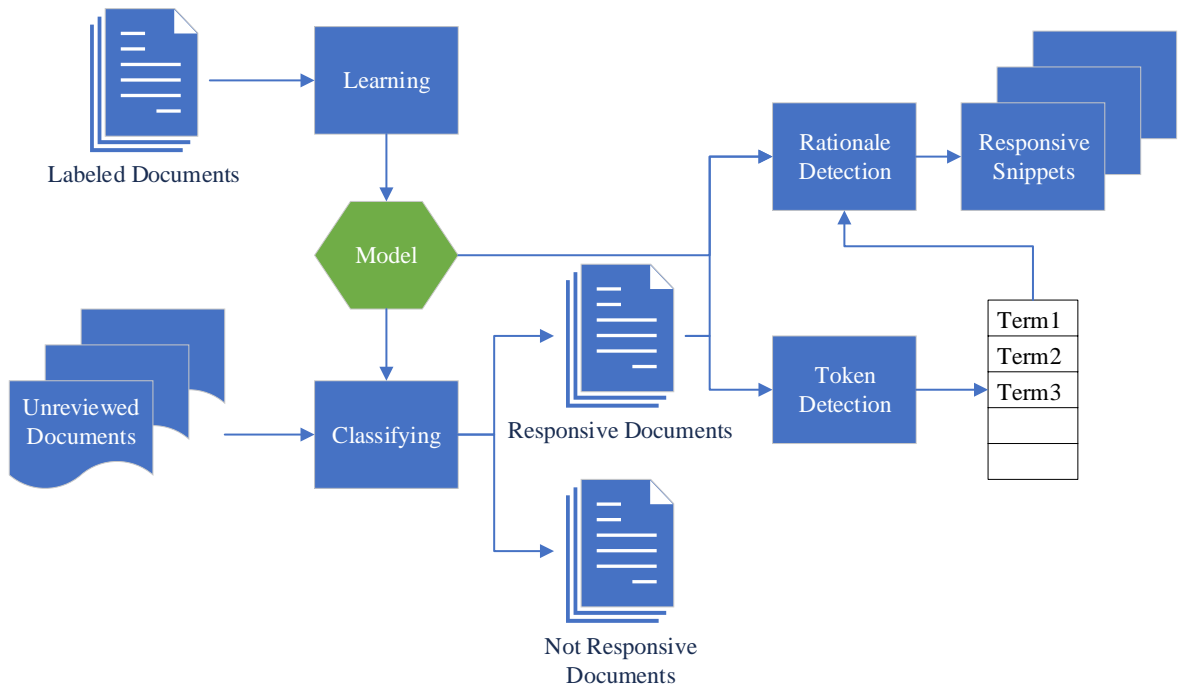

Figure 1. Framework for Explainable Predictive Coding

| |
|---|
| 4. Assign a score to all text snippets of responsive documents. |
| 5. The text snippet with the largest score is considered as a rationale for the document or all text snippets with scores larger than a threshold are considered as the rationales. |

The first component is called the <u>Snippet Score</u>, and we used a set of training documents, each of which was labeled as either responsive or not responsive by attorneys, and also a set of documents requiring classification. As mentioned above, this component contains of two phases. In the first phase, as is typical in traditional predictive coding, a model is generated using the set of training documents and the model is then deployed to identify a set of potentially responsive documents, using the model's predictions. In the second phase, the same model is applied to generate one or more rationales for each of the identified responsive documents.

To generate the rationales for a responsive document, we break the document into a set of small overlapping text snippets. We then apply the document model to classify these snippets, on the spectrum of highly likely to be responsive, down to highly likely to be not responsive. As such, the model assigns a probability score between 0 and 1 to each text snippet. The text snippets with the scores larger than a threshold or the snippets with the largest scores are selected as the rationales for the classifications.

Determining the optimal size of the text snippets is an open task because the size of the rationales is unknown in advance of generating the rationale. One approach to solve this problem is to use an iterative approach to break documents into a set of overlapping text snippets, as follows: first break a document into relatively large snippets – for snippets receiving large probability scores, continue to break them into smaller sizes. This process continues until probability scores stop increasing. In this paper, we use various predefined snippet lengths as a simplified version of this approach.

Similar to the first component, the model used to score text snippets in the second component is the document model. Instead of scoring text snippets directly, the model scores the document's remaining text with the snippet removed. We refer to this component as the <u>Snippet Complement Score</u>. As with the first component, the document is first broken into a set of small overlapping text snippets. For each snippet, we remove it from the document and then apply the document model to score the document with the snippet removed. The Snippet Complement Score is computed as follows:

$$\text{Snippet Complement Score}(s) = \begin{cases} 0, & DS(d) < DS(d-s) \\ (DS(d)-DS(d-s))/(DS(d)), & \text{Otherwise} \end{cases} \quad (1)$$

where $d$ is the document, $s$ is a snippet of the document $d$, $d - s$ is the document $d$ with the snippet $s$ removed, $DS(s)$ is the score of the snippet s, $DS(d)$ is the score assigned to the document $d$ by the document model, and $DS(d - s)$ is the score assigned to the document $d$ with $s$ removed by the document model. The Snippet Complement Score is measured as the normalized score reduction by removing the snippet from the document. The idea behind this scoring measure is that the removal of a highly responsive snippet from a responsive document should cause a reduction in the score when compared to the document's overall score. The more responsive the snippet is, the larger reduction in score. Text snippets with complement scores larger than a threshold are considered as rationales.

A linear classifier, such as SVM or logistic regression, establishes weights for tokens when generating a predictive model. Tokens with positive weights are considered to be responsive tokens. Considering this, we thought that we could use responsive tokens to identify rationales. This approach is inspired by several existing work. For example, [13] reported an approach to identify important words or n-grams for a prediction using a deep learning algorithm. Martens and Provost [10] proposed an algorithm-agnostic method, which finds the minimal set of words that change the classification if they are removed.

For this approach, we used a simple method based on a linear classifier and a bag of words representation of the documents to implement responsive token-based rationale detection. For each model, we generate a dictionary of important tokens for responsive documents. A token is added to the dictionary if its weight in the linear model is positive and larger than a threshold. Each token in the dictionary is associated with its weight in the linear model. For each document classified as responsive, we compute a contribution score for each of the tokens in the document and in the dictionary. The contribution score of a token is computed as follows:

$$\text{CScore}(t) = \text{weight}(t) \times \text{value}(t), \quad (2)$$

where $t$ is a token, $CScore(t)$ is the contribution score of $t$, $weight(t)$ is the weight of the token in the linear model, and $value(t)$ is the value of $t$ in representing the document. The value of a token could be binary, its frequency, normalized frequency, or tfidf value. The $n$ tokens with the highest contribution scores are considered to be responsive tokens. Using a linear classifier, the minimum set of tokens – those tokens considered the most important to explain the classification decision – can be found by continuingly adding the token with the largest contribution score to the minimum set until the model's classification changes the class to another (in this context from "responsive" to "not responsive"). This component is called the Keyword Score.

The <u>Snippet Token Score</u> of a text snippet is the sum of the contribution score of all tokens hit by the snippet. Specifically, the keyword score of a text snippet is computed as follows:

$$\text{SnippetTokenScore}(s) = \sum_{t \in s} \text{CScore}(t) / \sum_{t \in d} \text{CScore}(t), \quad (3)$$

where $t$ is a token, $s$ is a text snippet, and $d$ is the document. The contribution of keywords in the snippet is normalized by the total score of all the keywords in the document.

The three components described above could be used independently as approaches to identify rationales but they could also work together to identify rationales. We propose two simple methods, score-based and rank-based, to integrate these three components. In the <u>Score-Based Integration</u> method, the snippet score is a weighted linear sum of the scores of the three components:

SnippetFinalScore(s) = $w_1 \times$ SnippetScore($s$)
+ $w_2 \times$ SnippetComplementScore($s$)
+ $w_3 \times$ SnippetTokenScore($s$),

where SnippetScore($s$) is the snippet score generated by the document model, SnippetComplementScore($s$) is the score computed using formula (1), and SnippetTokenScore($s$) is the score computed using formula (3). $w_i$ ($i = 1, 2, 3$) are numeric weights, which can be adjusted.

Because the same document model is used to compute each component's score, they are not orthogonal. To accommodate the scale differences across the score types, we propose a <u>Rank-Based Integration</u> method, which derives the score of a component using a ranking of the documents based on the corresponding scores. To convert each of the above three types of scores to a ranked based score, we use the following formula:

$$\text{RS}_{\text{score}}(s) = 1/(k + \text{rank}_{\text{score}}(s))$$

, where $k$ is a constant set to 60, and $\text{rank}_{\text{score}}(s)$ is the rank of the snippet according to the score in descending order. This approach has been shown to be effective in merging results scored by different methods and setting $k$ to 60 is generally optimal according to previous research [14]. Under this Rank-Based integration method, the snippet score is a weighted linear sum of the rank scores of the three components:

SnippetFinalScore(s) = $w_1 \times \text{RS}_{\text{SnippetScore}}(s)$
+ $w_2 \times \text{RS}_{\text{SnippetComplementScore}}(s)$
+ $w_3 \times \text{RS}_{\text{SnippetTokenScore}}(s)$,

where $w_i$ (i = 1, 2, 3) are numeric weights that can be adjusted.

## IV. EXPERIMENTS

In this section, we report our results on a large data set from a real legal document review engagement. We describe the data set and the design of the experiments in Sections IV.A and IV.B, respectively. Results are reported in Section IV.C.

### A. The Data Set

The data set was collected from a legal matter that has since concluded. The data set contains documents including emails, Microsoft Office documents, PDFs, and other text-based documents. It consists of 688,294 documents manually coded by attorneys as responsive or not responsive. Among the 688,294 documents, 41,739 are responsive and the rest are not responsive. For each of the responsive documents, a rationale was annotated by a review attorney as the justification for coding the document as responsive. In practical terms, most rationales are continuous words, phrases, sentences or sections from the reviewed and labeled documents. A few rationales contain words that are comments from the attorney and do not occur in the documents. Some rationales may consist of more than one text snippet, which occur in different parts of the document.

Annotated rationales have a mean length of 52 words, with a standard deviation of 112.5 words. 97.5% of these rationales have fewer than 250 words. To reduce the effect of outliers, such as very long or very short rationales, we limit our rationales to those with ten or more words but fewer than 250 words and those that can be precisely identified in the data set – resulting in 23,791 responsive documents with annotated rationales in our population. These 23,791 documents established our responsive population, covering 57% of all the responsive documents in the manually coded population of 688,294 documents. Proportionally, we randomly selected 365,742 documents from the not responsive documents within the 688,294 population to define the not responsive population in our experiments.

### B. Experiment Design

The purpose of these experiments was to study the feasibility of the proposed approaches for automatically identifying rationales for responsive documents. The experiments simulate a real legal application scenario. In these experiments, a document model was trained using documents with existing responsive and not responsive labels from review by attorneys. The model was applied to classify all documents into responsive or not responsive. Then, the three approaches, based on individual components of the framework and the integrated approaches described in Section III, were applied to the documents classified as responsive to identify rationales that "explain" the models' responsive decision. In these experiments, a responsive classified document was divided into a set of overlapping snippets and each of the snippets was a candidate rationale. The snippet size $n$ = 50, 100, and 200, respectively with $n/2$ words overlapping between neighboring text snippets. Next, the rationale detection methods were applied to these snippets to determine if the snippets were rationales or not. A snippet was a true rationale if it overlapped with the annotated rationale with more than words, *(max(n,m))/2*, where *n* is the size of the snippet and *m* is the size of the annotated rationale.

The machine learning algorithm used to generate the models was Logistic Regression. Our prior studies demonstrated that predictive models generated with Logistic Regression perform very well on legal matter documents [15, 16]. Other parameters used for modeling were bag-of-words with 1-gram and normalized frequency [15]. The results reported in the next section were averaged over a five-fold cross validation.

*C. Results of the Experiments*

Table 2 details the statistics of the text snippets for each snippet size. *M* (1, 2, 3, 4, 5) top scoring snippets were selected as the identified rationales for each model. An identified snippet is a true rationale if it has an overlap span with the annotated rationale identified by the attorney reviewer that is larger than 50% of the smaller of either:

- the snippet window size, or;
- the annotated rationale size.

We believe an overlap of this type is a convincing criteria of algorithm performance to identify rationales in documents.

Table 2: Text Snippet Statistics

| Snippet Size | Total Number of Snippets | Number of Documents | Average Number of Snippets per document |
|---|---|---|---|
| 50 | 933,997 | 23,971 | 39 |
| 100 | 473,181 | 23,971 | 20 |
| 200 | 242,578 | 23,971 | 10 |

We conducted experiments using the three individual snippet detection approaches of the framework introduced in Section III and the two integration methods to identify rationales. In these experiments, for the two integration methods, we did a simple grid search to determine the optimal values of weights $w_1$, $w_2$, and $w_3$. Based on this search, we used $w_1 = 0.7$, $w_2 = 0.2$ and $w_3 = 0.1$ for results of both the scored-based and ranked based integration methods. For the Keyword Score method, we chose the top 100 keywords with the highest weights in our model. The total number of model keywords was 20,000.

The first set of experiments simulated a real legal matter scenario, namely identifying rationales for documents that were classified as responsive by the document model. As in most legal document review engagements, we set the cut-off score for the model to identify 75% of the all responsive documents. Table 3 reports the rationale recalls for different methods. For snippets with 50 words, the two integration methods performed the best, but only slightly better than the Snippet Score method, which is slightly better than the Snippet Complement Score method. The Keyword Score method performed the worst of the five approaches. For snippets with 100 words, Score-Based Integration performed the best for the top one, two and three snippets, while the Snippet Score method achieved higher recall for the top four, and five snippets. For snippets with 200 words, the Snippet Score method achieved the best performance for top two, three and four snippets, while the integration methods performed better for top one and five snippets.

Table 3: Rationale Recall (%) for documents identified as responsive (75% documents with the highest scores)

| Number of words in Snippet | Top K Snippets | Snippet Score | Snippet Complement Score | Keyword Score (100 keywords) | Score-Based Integration | Rank-Based Integration |
|---|---|---|---|---|---|---|
| 50 | 1 | 33.49 | 33.16 | 16.31 | **34.02** | 33.93 |
| | 2 | 46.23 | 43.73 | 28.97 | **46.45** | 46.40 |
| | 3 | 55.64 | 52.43 | 39.38 | 55.84 | **55.88** |
| | 4 | 62.07 | 58.20 | 47.48 | **62.41** | 62.22 |
| | 5 | 67.19 | 62.69 | 54.38 | **67.48** | 67.31 |
| 100 | 1 | 42.16 | 41.25 | 22.97 | **42.54** | 42.20 |
| | 2 | 55.65 | 52.62 | 39.36 | **55.74** | 55.70 |
| | 3 | 65.68 | 62.32 | 51.37 | **65.81** | 65.36 |
| | 4 | **72.11** | 68.48 | 60.54 | 71.97 | 71.93 |
| | 5 | **76.67** | 72.95 | 67.59 | 76.62 | 76.54 |
| 200 | 1 | 53.82 | 54.24 | 33.51 | **54.41** | 53.74 |
| | 2 | **67.59** | 66.01 | 55.44 | 67.51 | 67.55 |
| | 3 | **77.16** | 74.97 | 68.12 | 77.04 | 77.03 |
| | 4 | **82.82** | 80.55 | 76.21 | 82.74 | 82.71 |
| | 5 | 86.65 | 84.62 | 81.61 | 86.59 | **86.68** |

We also conducted experiments to identify rationales for all responsive documents (100%). Table 4 shows the results for all responsive documents. In most of the cases, either Scored-based or Rank-based Integration methods performed better than the methods based on individual scores, except that Snippet Score method performed best for top four and five snippets for snippets with 100 words and for top three and four snippets for snippets with 200 words. For all of the results, including results from Table 3, the performance of the two integration methods and the Snippet Score method are comparable, the Snippet Complement Score method did not perform as well as the above three approaches. In part, this could be due to the fact that a responsive document may include more than one responsive snippet. Again, the Keyword Score method performed significantly worse than the other approaches. This may be because we only selected a small portion (0.5%) of all model keywords, many of the snippets received scores of zero, which limited the classification potential of this method. We plan to experiment with selecting a larger portion of top weighted keywords as well as models of fewer keywords in future studies.

rationales were successfully identified by the individual approaches for the 75% recall set (we omitted results on the 100% recall set since both showed a similar pattern), measured by the Jaccard Index statistic. We can see the overlapping degree between the sets identified by the different individual approaches is low when the number of words in the snippets is small and when only the top one or two snippets are considered. This implies that it may be promising to integrate the individual components to deliver a more effective categorizing rationale identification solution.

Table 4: Rationale Recall (%) for all responsive documents

| Number of words in Snippet | Top K Snippets | Snippet Score | Snippet Complement Score | Keyword Score (100 keywords) | Score-Based Integration | Rank-Based Integration |
|---|---|---|---|---|---|---|
| 50 | 1 | 35.00 | 34.33 | 18.20 | **35.43** | 35.29 |
| | 2 | 48.19 | 45.37 | 31.99 | 48.08 | **48.20** |
| | 3 | 57.38 | 54.21 | 42.60 | 57.43 | **57.52** |
| | 4 | 63.69 | 60.10 | 50.73 | **63.88** | 63.78 |
| | 5 | 68.74 | 64.53 | 57.37 | **68.86** | 68.77 |
| 100 | 1 | 43.19 | 42.47 | 25.02 | **43.59** | 43.18 |
| | 2 | 57.19 | 54.52 | 42.42 | 57.14 | **57.24** |
| | 3 | 67.15 | 64.25 | 54.63 | **67.24** | 66.91 |
| | 4 | **73.65** | 70.39 | 63.38 | 73.50 | 73.49 |
| | 5 | **78.12** | 74.87 | 70.05 | 77.95 | 77.95 |
| 200 | 1 | 54.13 | 54.86 | 36.32 | **54.91** | 54.02 |
| | 2 | 69.21 | 67.89 | 58.75 | **69.24** | 69.19 |
| | 3 | **78.77** | 76.88 | 71.01 | 78.67 | 78.68 |
| | 4 | **84.17** | 82.23 | 78.46 | 84.07 | 84.07 |
| | 5 | 87.70 | 85.93 | 83.45 | 87.64 | **87.73** |

Table 5: Jaccard Similarity between sets of documents of which rationales were successfully identified by the individual approaches (75% documents with the highest scores)

| Number of words in Snippet | Top K Snippets | Snippet Score vs. Snippet Complement Score | Snippet Score vs. Keyword Score | Snippet Complement Score vs. Keyword Score |
|---|---|---|---|---|
| 50 | 1 | 59.27% | 23.84% | 23.12% |
| | 2 | 66.78% | 36.46% | 36.20% |
| | 3 | 70.04% | 46.47% | 45.38% |
| | 4 | 73.00% | 54.59% | 53.40% |
| | 5 | 75.79% | 61.17% | 59.87% |
| 100 | 1 | 69.19% | 29.52% | 29.24% |
| | 2 | 75.69% | 45.98% | 45.62% |
| | 3 | 78.08% | 57.00% | 56.22% |
| | 4 | 81.30% | 65.32% | 64.40% |
| | 5 | 83.25% | 71.59% | 70.51% |
| 200 | 1 | 79.63% | 40.60% | 40.52% |
| | 2 | 85.09% | 61.30% | 61.15% |
| | 3 | 86.77% | 72.22% | 71.58% |
| | 4 | 89.08% | 79.40% | 78.69% |
| | 5 | 91.45% | 83.95% | 83.67% |

In many cases of the two sets of experiments, we observed that the integration methods performed better than the individual component approaches, especially when only the top one or two snippets were considered or when the snippet window size was small, e.g. 50 words. We tried to better understand this observation by analyzing the overlap between the sets of documents of which rationales were successfully identified by the individual component approaches. Table 5 reports the similarity between the set of documents of which

V. ADDITIONAL OBSERVATIONS

While the results of our experiments provide effective methods to help "explain" a predictive model's classification decision, we also found that this approach can be very effective at identifying a predictive model's false negatives. This group of collaborators attempted to find documents with high text snippet scores (rationale scores) based on the document model but where the document model classified the overall document as not responsive. When applied to Jeb Bush's publicly available gubernatorial emails [17], we were able to successfully identify documents that were actually responsive (false negatives) but were otherwise classified as not responsive by the predictive model. For example, the score of one of the documents from this data set was 34.82. Choosing a cut-off score at 50 to define the boundary between responsive and not responsive classes, this document would be coded as not responsive by the predictive coding model. However, a snippet from this low scoring document (see Figure 2) received a score of 86.95, indicating it's likely responsive to the issue at hand: "Manatee Protection" – our model was created to target emails that discuss manatee protection. Further, excluding this text snippet from the document's overall text and rescoring the document with our model resulted in a score of: 2.92, a very low score indicating the remaining text is not responsive. After reading the full document's text we confirmed that the document was indeed about "Manatee Protection". The text snippet in Figure 2 is part of the original email in a chain of emails that make up this document. After this initial discussion of manatees, the rest of the email thread shifted to a long discussion about administrative topics, which prevented the document from receiving a higher score by the model. This example demonstrates that our approach can be used to identify responsive documents that are classified as not responsive incorrectly because a document contains multiple topics discussed at length.

```
- High-speed boat races in manatee
habitat are inappropriate.
- The race and spectator boats will
place the endangered Florida manatee
at an unacceptable risk for boat-
related harassment, injury and death.
- Manatee use of the area is well
documented and occurs year round.
- The Manatee population in Southwest
Florida is in decline and additional
threats like a high-speed boat race
in important manatee habitat must not
occur.
- The adverse effects of a high-speed
boat race on the Florida manatee are
not allowed by the ESA and MMPA and
result in unauthorized illegal "take"
of manatees.
-  The appropriate location for an
"offshore" race is off shore and
outside Tampa Bay!

Denise Nassif
Membership Services Representative

To learn about our programs
protecting endangered manatees please
visit our website at:
www.savethemanatee.org
```

Figure 2. Text Snippet Jeb Bush Gubernatorial Emails

We observed similar results on more than one confidential and real legal matter too. We suspect that there are documents across many real-world data sets with a short responsive statement that is overshadowed by its not responsive content – leading to misclassification by the model. This new and practical use case of targeting false negatives provides an exciting opportunity to help improve the quality of a predictive model's results in a legal document review. We plan to perform additional experiments to further explore this new use case.

VI. CONCLUSION

The authors believe that explainable predictive coding (text classification) has potential to dramatically advance the state of predictive coding and help significantly reduce document review costs. This paper proposed a framework for explainable text classification in legal document review. Specifically, we proposed three approaches based on individual components of the framework and two integrated methods of those components. We also conducted experiments on one large data set to compare these proposed approaches. Our experimental results demonstrated that the framework and integrated methods successfully target document explanations. In one scenario, explanations for at least one third of the responsive documents could be correctly identified. This means that for at least one third of all documents to be reviewed, contract lawyers would only need to review the 50-word snippets to quickly determine that those documents were responsive.

While the results reported in this case study just begin to scratch the surface of explainable predictive coding in the legal space, they are promising. The results also demonstrated that it is feasible to build text classification models to identify rationales automatically without using annotated rationales. These results open the door to conduct future studies in explainable predictive coding. We were excited to find that this framework can be used to target false negatives. Historically, in the legal domain, it has been challenging to address the predictive model's error and target false negatives. We plan to conduct more experiments to develop an effective and repeatable methodology to targeting a predictive model's false negatives.